\documentclass[epj]{svjour}

\usepackage{graphicx}
\usepackage{fancyhdr}
\def\makeheadbox{{
 }}
\def\mytitle{Lowest Landau Level of Relativistic Field Theories in a Strong Background Field} 
\def\myauthors{X. Calmet and M. Kober}  
\def\mytype{Parallel}
\def\mysession{Alternatives}

\begin{document}
\title{Lowest Landau Level of Relativistic Field Theories in a Strong Background Field}
\author{Xavier Calmet\inst{1}
\thanks{\emph{Email: xavier.calmet@ulb.ac.be}}%
 \and
 Martin Kober\inst{2}
}                     
%
%
\institute{Universit\'e Libre de Bruxelles, 
Service de Physique Th\'eorique, CP225,
Boulevard du Triomphe  (Campus plaine),
B-1050 Brussels, Belgium.
\and Institut f\"ur Theoretische Physik,
Johann Wolfgang Goethe-Universit\"at,
Max-von-Laue-Stra\ss e 1,
D-60438 Frankfurt am Main, Germany.}
%
\date{}
\abstract{
We consider gauge theories in a strong external magnetic like field. This situation can appear either in  conventional four-dimensional theories, but also naturally in extra-dimensional theories and especially in brane world models. We show that in the lowest Landau level approximation, some of the coordinates become non-commutative. We find physical reasons to formal problems with non-commutative gauge theories such as the issue with SU(N) gauge symmetries. Our construction is applied to a minimal extension of the  standard model. It is shown that the Higgs sector might be non-commutative whereas the remaining sectors of the standard model remain commutative. Signatures of this model at the LHC are discussed. We then discuss an application to a dark matter sector coupled to the Higgs sector of the standard model and show that here again, dark matter could be non-commutative, the standard model fields remaining commutative. 
} 
\maketitle
Gauge theories formulated on non-commutative \linebreak spaces have received a lot of attention over the last decade. The main reason is that they were discovered to appear in a certain limit of string theory\cite{Connes:1997cr,Seiberg:1999vs,Schomerus:1999ug}. Non-commuting coordinates do appear generically whenever one studies a physical system in an external background field in the first Landau level approximation. This phenomenon was discovered by Landau in 1930 \cite{Landau}. A textbook example is an electron in a strong magnetic field. In the framework of string theory, the effective low energy four dimensional theory describing strings ending on a brane in the presence of a strong external background field is shown to be non-commutative\cite{Connes:1997cr,Seiberg:1999vs,Schomerus:1999ug}.

It is notoriously difficult to construct a  \linebreak  non-commutative version of the standard model. There are different approaches in the literature \cite{Calmet:2001na,Chaichian:2004yw}. The main difficulty is to obtain the right gauge symmetries i.e. SU(N) groups necessary to describe the standard model. The issue here is that the commutator of two non-commutative Lie algebra valued gauge transformations is not a gauge transformation unless one chooses U(N) Yang-Mills symmetries and the fundamental, anti-fundamental or adjoint representations. This no-go theorem can be avoided if one considers the enveloping algebra. However this may not seem very natural. This is the motivation for the present work. We shall study relativistic field theories in a strong external potential to identify the physical reason for this technical problem. We shall start from classical gauge theories formulated on a regular space-time which are perfectly well behaved and renormalizable theories and consider them in a strong external field, then we shall consider their first Landau Level. Different field theories have been considered in the lowest Landau level approximation leading to more or less exotic non-commutative gauge theories, see e.g. \cite{Jackiw:2001dj} or \cite{Horvathy:2002wc}. However, we wish to consider physical situations which lead to the kind of non-commutative gauge theories found in \cite{Connes:1997cr,Seiberg:1999vs,Schomerus:1999ug} to understand the physical reason for some of the pathologies of these theories. Furthermore our construction allows us to consider models where only certain sectors of the theory are noncommutative.

Let us first consider a charged scalar field in a strong magnetic field. The action is given by
\begin{eqnarray} \label{action1}
S&=& \int d^4 x \Big ( (\bar D_\mu \phi)^*  (\bar D^\mu \phi) - V(\phi^*\phi)
\\ \nonumber &&  -\frac{1}{4} \bar F_{\mu\nu} \bar F^{\mu\nu} \Big ),
\end{eqnarray}
where $\bar D_\mu=\partial_\mu + i q {\cal A}_\mu$ and $\bar F_{\mu\nu} =-i[\bar D_\mu,\bar D_\nu]$. This theory is gauge invariant under U(1) gauge transformations and is renormalizable. Let us now study this theory in the limit of a strong external magnetic field. We consider quantum fluctuations $A_\mu$ around $C_\mu$ which is the background field which corresponds to the constant magnetic field. We then have $\bar D_\mu=\partial_\mu + i q A_\mu + i q C_\mu = D_\mu +  i q C_\mu$ and the action becomes
\begin{eqnarray} \label{action2}
S&=& \int d^4 x   \Big ( (D_\mu \phi)^*  (D^\mu \phi) - V(\phi^*\phi)  \\ \nonumber &&
 - i q \phi^*  C_\mu D^\mu \phi + i q (D_\mu \phi)^* C^\mu \phi
+  q^2 \phi^* C_\mu C^\mu \phi \\ \nonumber  &&
 -\frac{1}{4}  F_{\mu\nu}  F^{\mu\nu} -\frac{1}{4}  C_{\mu\nu}  C^{\mu\nu}-\frac{1}{4}  F_{\mu\nu}  C^{\mu\nu}
 \\ \nonumber && -\frac{1}{4}  C_{\mu\nu}  F^{\mu\nu} \Big),
\end{eqnarray}
where $C_{\mu\nu}=\partial_\mu C_\nu - \partial_\nu C_\mu$. To be more specific we shall pick $C_\mu= (0, \frac{ B y}{2},  -\frac{B x}{2}, 0)$ which leads to a constant magnetic field of magnitude $B$ in the $z$-direction. Note that a strong external magnetic like field does not imply that the quantum fluctuation is strongly coupled to the scalar field. The action (\ref{action2}) has a remaining gauge invariance: $\delta \phi=i\alpha \phi$, $\delta A_\mu= \partial_\mu \alpha$  and the background field is kept invariant.
The classical canonical momenta of the center of mass of particle $\phi$ is given by 
$\pi_\mu=p_\mu+  q A_\mu + q C_\mu$.

The first quantization of the classical Hamiltonian implies that the coordinates and the spatial components of the canonical momentum  do not commute:
\begin{eqnarray}
[x^i,\pi^j]=i \hbar \delta^{ij}.
\end{eqnarray}
We now express the canonical momentum in terms of the kinematical one and find
$[x^i, p^j+ q A^j + q B  \epsilon^{jk}x_k]=i \hbar \delta^{ij}$ for $i,j \in \{1,2\}$. Let us now consider the limit $\sqrt{B} \gg m$ and $|C^\mu| \gg |A^\mu|$. In this limit the terms involving the kinematical momentum $p^j$ and the potential $A^j$ can be neglected:
\begin{eqnarray} \label{ncalg}
[x^i, x^j]= i \hbar \frac{1}{qB}\epsilon^{ij}\equiv i \theta^{ij}.
\end{eqnarray}
This means that the scalar field is non-commutative in the $x-y$ plane. It should be noted that our result is not a gauge artifact, the very same result would be obtained if we had chosen e.g. the Landau gauge $C_\mu= (0, B y,0, 0)$.
Furthermore, it is easy to see that since Lorentz covariance is explicitly broken by the background field new Lorentz violating vertices involving the gauge boson $A_\mu$ will be generated through its interaction with the background field. In particular three gauge bosons and four gauge bosons vertices which are typical of non-commutative gauge theories are generated. We have just shown that in the limit $m\ll \sqrt{B}$, the coordinates $x$ and $y$ of the scalar field  do not commute, let us rename them $\hat x$ and $\hat y$. In the limit $m\ll \sqrt{B}$, local gauge transformations of the scalar field involve non-commuting coordinates: $\delta_\alpha \phi = i \alpha(t,\hat x, \hat y, z) \phi(t,\hat x, \hat y, z)$, in order to build a gauge invariant action, the gauge boson has to transform according to 
$\delta_\alpha A_\mu(\hat x)=\partial_\mu \alpha (\hat x)+ i [\alpha(\hat x),A_\mu(\hat x)]$.
The low energy action is then given by
\begin{eqnarray} \label{action3}
S&=& \int d^4 x  \big ((D_\mu \phi(\hat x))^*  (D^\mu \phi(\hat x)) - 
V(\phi(\hat x)^*\phi(\hat x)) \\ \nonumber &&
  -\frac{1}{4}  F_{\mu\nu}(\hat x) F^{\mu\nu}(\hat x) \big),
\end{eqnarray}
with $F_{\mu\nu}=-i[D_\mu(\hat x),D_\nu(\hat x)]$. 
Using the Weyl quantization procedure, it is easy to replace the \linebreak  non-commuting coordinates in the argument of the field $\phi$ by commuting ones 
\begin{eqnarray} \label{action4}
S&=& \int d^4 x \Big ((D_\mu \phi)^* \star (D^\mu\phi) - V(\phi^* \star  \phi) 
\\ \nonumber && -\frac{1}{4}  F_{\mu\nu} \star  F^{\mu\nu}  \Big),
\end{eqnarray}
where the star product is given by
$f \star g = f  e^{i \partial_i  \theta^{ij}  \partial_j} g$
with $\theta^{ij}=\frac{\hbar}{qB} \epsilon^{ij}$ for $i,j \in \{1,2\}$ and $\theta^{\mu\nu}=0$ in the time and $z$-directions.

It should be noted to our derivation that it is not specific to a scalar field theory since the important point comes from the equations of motion which are the Klein-Gordon equations. Since every component of a spinor field satisfies the Klein-Gordon equations, our result applies to spinor field as well. Our first result is that the action (\ref{action2}) is very identical to a U(1) non-commutative gauge theory with a non-commutativity in the $x-y$ plane. We find that a non-commutative gauge theory is very closely related to a commutative gauge theory in a strong external field in the limit that the mass of the particle is small compared to the external background field. It is well known that the action we started from is well behaved at the quantum level and in particular that it is renormalizable. On the other hand the non-commutative action (\ref{action4}) is not renormalizable and suffers from UV/IR mixing. This is a strong hint that the issues with the quantum field calculations involving the action (\ref{action4}) should disappear in the limit where more and more Landau levels are included in the calculations. However, we should point out that the naive limit $B\to 0$ which would correspond to a vanishing external field implies an infinite non-commutative parameter. The limits $B\to 0$ and $\sqrt{B}\gg m$ do not commute. This is clearly another kind of UV/IR mixing and probably the origin of UV/IR mixing in the quantized version of the theory.

We can now push our analysis further and consider Yang-Mills theories instead of a simple U(1) theory. To be very concrete let us consider a SU(2) Yang-Mills theory. In that case there are three gauge potentials $B^1_\mu$, $B^2_\mu$ and  $B^3_\mu$. We see that the same procedure as the one outlined in this work leads to two canonical momenta, one for each of the components of the doublet $\vec \phi=(\phi_1,\phi_2)$. To be very precise let us consider the canonical momentum $\pi^i_1$ of the particle described by the field $\phi_1$ and $\pi^i_2$ which corresponds to the particle  $\phi_2$. It is clear that it only depends on $B^{3}_\mu$ since the generator $T^3$ is the only diagonal one. However $T^1$ and $T^2$ are not diagonal and thus  $B^{1}_\mu$ and $B^{2}_\mu$  do not contribute to the canonical momenta. One finds
$\pi^i_1=p^i_1+  g B^{3 i} + g D^{3 i}$, where $B^3_i$ is the fluctuation around the strong external field $D^3_i$ and $\pi^i_2=p^i_2-  g B^{3 i} - g D^{3 i}$. Let us now assume that the non-vanishing components of the strong external field $D^3_i$ are given by $E  \epsilon_{ij}x^j$, we find 
$[x_i, p_j+ g B_j^3 + g E  \epsilon_{jk}x^k]=i \hbar \delta_{ij}$ and 
$[x_i, p_j-g B_j^3 - g E \epsilon_{jk}x^k]=i \hbar \delta_{ij}$
for $i,j \in \{1,2\}$ let us now consider the limit $\sqrt{E} \gg m$ and $|D^3_j| \gg |B_j^3|$ one finds
$[x^i, x^j] = i \hbar \frac{1}{g E}\epsilon^{ij}$ and simultaneously $[x^i, x^j] = - i \hbar \frac{1}{g E}\epsilon^{ij}$ which is clearly inconsistent. In other words, there  is no non-commutative SU(2) theory equivalent to a SU(2) gauge theory restricted to its first Landau Level. However if we had started from a U(N) gauge group, one of the generators would be proportional to the identity matrix and we could have chosen the strong external field in the direction of the identity matrix and obtained a consistent non-commutative algebra. In that case the first Landau level of a gauge theory can be described in terms of a dual non-commutative gauge theory as long as the external strong field is chosen in the direction of the identity matrix. This is the physical origin of the formal problem with SU(N) gauge invariance mentioned at the beginning of this work.

Furthermore, it should be stressed that the commutative action (\ref{action2}) is not gauge invariant under regular gauge transformations for U(N) (N$>$1) gauge groups unless the background field transforms as well: $\delta A^\mu= \partial^\mu \alpha + i [\alpha, A^\mu]$ and $\delta C^\mu=  i [\alpha, C^\mu]$.
Note that we are using a different convention than in the background quantization technique where the background field transforms as a gauge field whereas the quantum fluctuation transforms homogeneously \cite{Weinberg:1996kr}. The subtlety only appears for N$>$1. This suggests a generalization of non-commutative gauge transformations to
\begin{eqnarray}
\delta A^\mu&= &\partial^\mu \alpha + i \alpha \star A^\mu  - i A^\mu  \star \alpha \\ 
\delta C^\mu&=&   i \alpha \star C^\mu  - i C^\mu  \star \alpha,
\end{eqnarray}
where we set $g=1$. It is also suggestive that the background field which is closely related to the non-commutative parameter through an equation such as eq. (\ref{ncalg}) should be introduced in the action:
\begin{eqnarray} \label{action5}
S&=& \int d^4 x \Big ((D_\mu \phi)^\dagger  \star (D^\mu \phi) - V(\phi^\dagger \star \phi) 
\\ \nonumber &&
 - i  \phi^\dagger \star  C_\mu \star   D^\mu \phi + i (D_\mu \phi)^\dagger \star  C^\mu \star \phi
 \\ \nonumber &&
+ \phi^\dagger \star  C_\mu \star   C^\mu \phi   
-\frac{1}{4}  F_{\mu\nu} \star  F^{\mu\nu} -\frac{1}{4}  C_{\mu\nu} \star  C^{\mu\nu}
\\ \nonumber &&
-\frac{1}{4}  F_{\mu\nu}  \star C^{\mu\nu}-\frac{1}{4}  C_{\mu\nu} \star  F^{\mu\nu} \Big).
\end{eqnarray}
In the sequel we shall however restrict our considerations to U(1) gauge theories where this subtlety is irrelevant. 

Let us now apply this idea to physics beyond the standard model. If the U(1) external field we are considering couples only to one specie of particle we would have a reason to explain why only a certain sector of the model is non-commutative. It is tempting to identify the scalar field we have introduced with the Higgs field of the standard model. However, the Higgs field of the standard model is charged under SU(2) $\times$ U(1) and this would lead to a SU(2) non-commutative theory which is as explained previously not consistent for fields which are Lie algebra valued. Furthermore, it is not possible to gauge the standard model Higgs doublet under a new U(1) without affecting its charge assignment under the standard model gauge group. However, there has been a growing interest \cite{Hill:1987ea,vanderBij:2006ne,Calmet:2002rf,Calmet:2003uj,Calmet:2006hs,Patt:2006fw,O'Connell:2006wi} for particles which are not charged under the gauge group of the standard model or almost decoupling from the action of the standard model. Furthermore, scalar singlets are interesting dark matter candidates \cite{McDonald:1993ex} and could explain why the Higgs boson of the standard model has not yet been discovered \cite{vanderBij:2006pg}. Let us consider the coupling of the action  (\ref{action1}) to the standard model and we assume that $\phi$ is a SU(3)$\times$SU(2)$\times$U(1)$_Y$ singlet, but that it is charged under a new U(1)$_E$ gauge group under which standard model particles are singlets. We shall call this new particle the e-photon.  Let us assume that the e-photon has a vacuum expectation which fills the universe which will single out a preferred direction in space-time. There are different model building options which will affect the precise form of the non-commutative tensor $\theta^{\mu\nu}$. For example, the e-photon and $\phi$ could for example be living in extra-dimensions and the standard model confined to a brane in which case the non-commutativity could be in three dimensions. If the new degrees of freedom are confined to  live in four dimensions then  we would have non-commutativity in only two-dimension in the plane perpendicular to the direction of the external strong field.

Let us consider the scalar sector of the theory. We have 
\begin{eqnarray} \label{model1}
S&=& \int d^4 x \Big ( (\bar D_\mu \phi)^*  (\bar D^\mu \phi) 
- m_\phi^2 \phi^*\phi   \\ \nonumber   && 
- \lambda_\phi (\phi^*\phi)^2 
+   ( D_\mu H)^\dagger  (D^\mu H)  \\ \nonumber   &&   - m_H^2 H^\dagger H - \lambda_H (H^\dagger H)^2  
 +  \lambda \phi^*\phi  H^\dagger H \Big),
\end{eqnarray}
where $H$ is the Higgs doublet of the standard model and $\phi$ is the new scalar singlet charged under the new U(1)$_E$ interaction. The SU(2)$\times$ U(1) symmetry of the standard model has to be spontaneously broken, i.e. the doublet acquires a vacuum expectation value and using the unitary gauge, one has $H=(0,h+v)$ where $v^2= - m_H^2/(2\lambda_H)$. However, we have two options for the U(1)$_E$ gauge symmetry. Let us first consider the case where the extra U(1) is not spontaneously broken, in other words $\phi$ does not acquire a vacuum expectation value. In that case the scalar potential is given by  $V[h,\phi \phi^*]= -2 m_H^2 h^2+\lambda_H (h+v)^4+ \lambda (h+v)^2 \phi \phi^*+m_\phi^2 \phi \phi^* + \lambda_\phi \phi \phi^*\phi \phi^*$.
It is easy to show that the e-photon and the usual photon do not mix. Furthermore there is no coupling between the e-photon and the fermions of the standard model. This new long range force is thus not in conflict with experiments. The new charged scalars are protected by the exact U(1)$_E$ symmetry and thus dark matter candidates. Furthermore, although the carrier of the new force in the dark matter sector are massless, the bounds on a fifth force in the dark matter sector \cite{Kesden:2006zb} do not apply to our model because the e-photon does not couple to regular matter. Let us now assume that the e-photon has some vacuum expectation value such that it correspond to a strong magnetic-type field in the $z$-direction and consider this model in the first Landau Level approximation. We find that the scalars $\phi$ have non-commuting coordinates in that limit. The non-commuting coordinates can be removed at the expense of introducing a star product $V[h,\phi \star \phi^*]= -2 m_H^2 h^2+\lambda_H (h+v)^4+ \lambda (h+v)^2 \phi \star \phi^* + m_\phi^2 \phi \star \phi^* + \lambda_\phi \phi \star \phi^* \star \phi \star \phi^*$,
i.e. the only non-commutative interaction are those which involve the field $\phi$ and obviously the e-photon. If its mass is low enough, this dark matter candidate could be produced at the LHC through the decomposition of a Higgs boson. The decay rate $\Gamma(Higgs \to \phi \phi^*)=  \frac{1}{4 \pi} \lambda^2 v^2 \frac{\sqrt{m_H^2- 4m_\phi^2}}{2 m_H^2}$ is basically commutative.
However, the self-interaction of the $\phi$-mesons and of the e-photon are non-commutative and the non-commutative nature of this sector could be checked by searching for the usual characteristic non-commutative self-interactions of the e-photon. The cross section for the dark matter candidate at the Tevatron and LHC corresponds to the one of a singlet added to the standard model in the commutative case and there should thus be a clear signal. However, detecting the non-commutative nature of the dark matter sector at a hadron collider will be a difficult task since one would have to search for the typical self-interactions of the e-photon. However, one would expect that the background field will impact the distribution of dark matter in our universe which would allow to identify a preferred direction in space-time.

Another option is to assume that the remaining U(1)$_E$ is spontaneously broken by a vacuum expectation value of  the $\phi$-mesons  in which case the e-photon acquires a mass. In that scenario, the $\phi$-mesons are not dark matter candidates. However, the residual degree of freedom after U(1)$_E$ symmetry breaking, which we call $\sigma$,  will mix with the standard model Higgs boson. One finds: $h_{\mbox{phys}}= \cos \alpha   \ h + \sin \alpha \ \sigma$, $\sigma_{\mbox{phys}}= \cos \alpha \ \sigma - \sin \alpha \ h$ where $\alpha$, the mixing angle, is determined by the scalar potential. When we consider the model in a strong external potential corresponding to a strong magnetic-like field in the $z$-direction and in the lowest Landau level limit, we find that both scalar fields are non-commuting in the $x-y$ plane whereas the remaining fields of the standard model are commutative. In that case the only sector of the theory which would exhibit a non-commutative nature is the scalar potential sector. The new non-commutative operators are $v_h h \star \phi \star \phi$, $v_\phi h \star h \star \phi$ and $h \star h \star \phi \star \phi$. Because of the trace property of the star product ($\int d^4x  f \star g= \int d^4x g \star f =\int d^4x f g$), the interactions of the two scalar degrees of freedom with the fermions and gauge bosons of the standard model are commutative.

{\it Conclusions:}  We have shown that the phenomenon discovered by Landau in 1930 appears in relativistic field theories. We find physical reasons to the  formal problems with non-commutative gauge theories such as the issue with SU(N) gauge symmetries. We apply our construction to a minimal extension of the  standard model and show that the Higgs sector might be non-commutative whereas the remaining  sectors of the standard model remain commutative. We discuss the signatures of this model at the LHC. We then discuss an application to a dark matter sector coupled to the Higgs sector of the standard model and show that here again, dark matter could be non-commutative, the standard model fields remaining commutative. 

{\it Acknowledgments:}
This work was supported in part by the IISN and the  Belgian science policy office (IAP V/27). 

%
%

\end{document}